\documentclass{osa-article}

\usepackage{amsmath}

\journal{osajournal}

\articletype{Research Article}

\usepackage{lineno}

\begin{document}

\title{Image-Plane Self-Calibration in Interferometry}

\author{Christopher L.Carilli,\authormark{1,*} Bojan Nikolic,\authormark{2} and Nithyanandan Thyagarajan\authormark{3} }

\address{\authormark{1}National Radio Astronomy Observatory, P. O. Box 0, Socorro, NM 87801, US\\
\authormark{2}Astrophysics Group, Cavendish Laboratory, University of Cambridge, Cambridge CB3 0HE, UK\\
\authormark{3}Commonwealth Scientific and Industrial Research Organisation (CSIRO), Space \& Astronomy, P. O. Box 1130, Bentley, WA 6102, Australia
}

\email{\authormark{*}ccarilli@nrao.edu} 



\begin{abstract}
  We develop a new process of image plane self-calibration for interferometric imaging data. The process is based on Shape-Orientation-Size (SOS) conservation for the principal triangle in an image generated from the three fringes made from a triad of  receiving elements, in situations where interferometric phase errors can be factorized into element-based terms.  The basis of the SOS conservation principle is that, for a 3-element array, the only possible image corruption due to an element-based phase screen is a tilt of the aperture plane, leading to a shift in the image plane. Thus, an image made from any 3-element interferometer represents a true image of the source brightness, modulo an unknown translation. Image plane self-calibration entails deriving the unknown translations for each triad image via cross-correlation of the observed triad image with a model image of the source brightness. After correcting for these independent shifts, and summing the aligned triad images, a good image of the source brightness is generated from the full array, recovering source structure at diffraction-limited resolution. The process is iterative, using improved source models based on previous iterations. We demonstrate the technique in the high signal-to-noise context, and include a configuration based on radio astronomical facilities, and simple models of double sources. We show that the process converges for the simple models considered, although convergence is slower than for aperture-plane self-calibration for large-$N$ arrays. As currently implemented, the process is most relevant for arrays with a small number of elements. More generally, the technique provides geometric insight into closure phase and the self-calibration process. The technique is generalizable to non-astronomical interferometric imaging applications across the electromagnetic spectrum.
\end{abstract}


\section{Introduction} \label{sec:intro}

Self-calibration is a well established technique for generating accurate, diffraction limited images of sources from interferometric data that may have large element-based phase distortions \cite{Cornwell+Wilkinson1981,  Cornwell+Fomalont1999, Cotton1995, Pearson1999, Pearson+Readhead1984, Readhead+Wilkinson1978, Schwab1981, Schwab1980, Wilkinson1989, Wilkinson+1979, Perley1999, Meimon+2008, Lannes1998b} (Note: the term element herein corresponds to one coherent voltage receiver in the aperture plane, such as an antenna and coherent amplifier in radio interferometry, or a siderostat in optical interferometry). It is well known that, in element-based self-calibration, the `closure phase', which is the phase of the triple product (commonly known as the `bispectrum') of interferometric visibilities (spatial coherence) on a closed triad of receiver elements, i.e. any 3-element array, is conserved (see Section~\ref{sec:SOS};
\cite{Jennison1958, Baldwin+1986, Cornwell1987, TMS2017, Tuthill+2000, Lohmann+1983, Buscher2015, Monnier2007b, Quirrenbach2001}). If propagation, or instrumental, phase corruptions are factorizable into element-based gain terms (often called the `piston phases' in optical interferometry; \cite{Meimon+2008}), the closure phase remains a true measurement of the source brightness properties, independent of the element-based errors. 

Self-calibration, and closure-phase, have been formulated in the context of measurements in the aperture plane of the instrument. We have recently shown how closure phase manifests itself in the Fourier conjugate image plane, via  Shape-Orientation-Size (SOS) conservation for the principal triangle in an image generated from the three fringes from any triad of array elements \cite{Thyagarajan+Carilli2022}. 

A natural consequence of SOS conservation is that every image generated from a triad of receiver elements of an interferometer represents a true image of the source, modulo an unknown overall translation due to the element-based phase errors. The translations are idiosyncratic to each triad, hence summing the uncalibrated triad images will not produce a coherent image of the source, meaning an image that recovers source structure to the diffraction-limited resolution. However, SOS conservation raises the interesting possibility of performing image-plane only self-calibration by finding and correcting for these independent triad translations. In this paper, we show that these translations can be derived through cross correlation of triad images with a simple, {\sl a priori} source brightness model, then subsequent iteration. The process is a direct analog to what is performed in aperture plane self-calibration, but is performed in the image plane, using images made from triads of antennas, not visibilities and element-based phases in the aperture plane. 

We present the justification and methodology of image plane self-calibration based on SOS conservation. We demonstrate that it works, at least for simple, but realistic, models of the source and array configuration with high signal to noise. While these demonstrations are done in the context of astronomical interferometry, the technique is generalizable to broader applications of interferometric calibration and imaging, and Fourier optics, across the electromagnetic spectrum \cite{Born+Wolf1999}.

We find the image plane self-calibration process, as implemented in our initial study, does not converge as quickly as aperture plane self-calibration for arrays with a large number of elements. Hence, while we demonstrate that the image plane technique works in the simple tests employed, applications remain to be explored where the technique is advantageous in practice. In Section~\ref{sec:discussion}, we identify the criteria under which the technique might be considered, particularly in the case of small N arrays.

\section{Process} \label{sec:process}

We briefly review some of the concepts applied herein. Details can be found in \cite{Thyagarajan+Carilli2022}, or specific references below. We then present the image plane self calibration process, and the specific models and algorithms employed in this initial validation of the technique. Again, our derivation is performed in the context of astronomical interferometry, but is generalizable to broader applications of interferometric imaging. 

\subsection{Self-calibration in Interferometry}\label{sec:calib}

The spatial coherence (or \textit{visibility}), $V_{ab}(\nu)$, of two voltages of the same polarization, measured by interferometers in the aperture plane at a quasi-monochromatic frequency, $\nu$, is related to the intensity distribution of the sky, $I(\hat{\mathbf{s}},\nu)$, as \cite{vanCittert34,Zernike38,Born+Wolf1999,TMS2017,Clark1999}:
\begin{align}
    V_{ab}(\nu) &= \int_\textrm{sky} A_{ab}(\hat{\mathbf{s}},\nu) I(\hat{\mathbf{s}},\nu) e^{-i2\pi \mathbf{u}_{ab}\cdot \hat{\mathbf{s}}} \mathrm{d}\Omega \, , \label{eqn:VCZ-theorem}
\end{align}
where, $a$ and $b$ denote a pair of array elements, $\hat{\boldsymbol{s}}$ denotes a unit vector in the direction of any location in the image, $A_{ab}(\hat{\mathbf{s}},\nu)$ is the spatial response (the `power pattern') of each element pair, $\mathbf{u}_{ab}=\mathbf{x}_{ab} (\nu/c)$ is referred to as the ``baseline'' vector which is the vector spacing ($\mathbf{x}_{ab}$) between the element pair expressed in units of wavelength, and $\mathrm{d}\Omega$ is the differential solid angle element on the image plane. 

The measurements at the receiving elements are, however, corrupted by distortions introduced by the propagation medium and receiver electronics that can be described as a multiplicative element-based complex gain factor, $G_a(\nu)$. Thus, the corrupted measurements are given by:
\begin{align}
    V_{ab}^\prime(\nu) &= G_a(\nu) \, V_{ab}(\nu) \, G_b^\star(\nu) \, , \label{eqn:uncal-vis}
\end{align}
where, $\star$ denotes a complex conjugation. 

The process of calibration determines these complex gain factors that correspond to the element-based distorting effects in Equation~(\ref{eqn:uncal-vis}). Calibration is typically done with one or more bright celestial objects (`calibrators'), whose visibilities are accurately known \cite{TMS2017,SIRA-II}. Equation~(\ref{eqn:uncal-vis}) is then inverted to derive the complex voltage gains, $G_a(\nu)$ \cite{Schwab1980, Schwab1981}. If these gains are stable over the calibration cycle time, they can then be applied to the visibility measurements of the target source, to obtain the true sky visibilities, and henceforth the sky brightness distribution, typically using Fourier transforms \cite{TMS2017,SIRA-II}.

Calibration can also be done using the target source itself, if it is sufficiently bright. This is called ``self-calibration'' \cite{Cornwell+Fomalont1999, Meimon+2008}. In self-calibration, an initial {\sl a priori} sky brightness model of the target object is used to predict the true visibilities. Again, Equation~(\ref{eqn:uncal-vis}) can then be inverted to determine the element-based calibration terms which are then applied to obtain an updated sky model. The process is iterated until convergence is achieved. This type of calibration is carried out using measurements in the aperture plane, namely, visibilities, and we hereafter refer to it as ``aperture plane self-calibration'' (APSC) \cite{Cornwell+Wilkinson1981, Cornwell+Fomalont1999, Cotton1995, Pearson+Readhead1984, Readhead+Wilkinson1978, Schwab1981, Schwab1980, Wilkinson1989}. 

\subsection{Closure Phase and Shape-Orientation-Size Conservation}\label{sec:SOS}

Closure phase is a quantity defined early in the history of astronomical interferometry, as a measurement of properties of the sky brightness that is robust to element-based phase corruptions \cite{Jennison1958}, and hence is useful in situations where phase calibration may be difficult. Closure phase is the sum of three visibility phases measured cyclically on three interferometer baseline vectors forming a closed triangle, i.e., closure phase is the argument of the bispectrum (product of three complex visbilities in a closed triad of elements):
\begin{align}
    \phi_{abc}(\nu) &= \phi_{ab}(\nu) + \phi_{bc}(\nu) + \phi_{ca}(\nu) \, . \label{eqn:CPhase}
\end{align}

The advantage in using closure phase is that it is independent of element-based corruptions \cite{Jennison1958,TMS2017,SIRA-II,Carilli+2018,Thyagarajan+2018, Monnier2007b}, because the antenna-based phase terms cancel, leaving the closure phase $\phi_{abc}^\prime(\nu)=\phi_{abc}(\nu)$, where, again, prime denotes measured, and un-primed true, sky brightness quantities, as per Eqn.~(\ref{eqn:uncal-vis}). 

The 1970s and early 1980s saw the use of the ``hybrid mapping'' technique \cite{Wilkinson+1977, Pearson+1981,Readhead+1988} that employed closure phase and closure amplitudes as the primary measurements against which to compare models. Starting with a simple model that included any \textit{a priori} knowledge of the morphology, the measurements of closure quantities were compared against the corresponding model which were then iteratively used to update the model until a convergence was achieved between the model and the measurements. This technique of using closure phases and visibility amplitudes directly to generate interferometric images has been revisited recently in the context of millimeter-wavelength Very Long Baseline Interferometry (VLBI) with the Event Horizon Telescope, for which a forward-modeling approach was employed to identify sky images most consistent with the interferometric measurements of closure phases and visibility amplitudes \cite{eht19-4}. 

The subsequent development of APSC \cite{Cornwell+Wilkinson1981, Cornwell+Fomalont1999, Cotton1995, Pearson+Readhead1984, Readhead+Wilkinson1978, Schwab1981, Schwab1980, Wilkinson1989} provided a more general and powerful solution to the calibration and imaging problem by using an initial model and adjusting only the element-based gains to match the model and measured visibilities.  As only multiplicative element-based corrections are applied, it was shown that APSC inherently preserves the closure quantities, which are invariant to the element-based contributions \cite{Jennison1958,TMS2017}. 

\cite{Cornwell1987} has shown the close mathematical relationship between closure phase and the triple product imaging technique to recover phase information and source structure at the diffraction limit in optical speckle masking \cite{Lohmann+1983,Weigelt+Wirnitzer1983}. 

As a counterpart to the well-known invariance of closure phase to element-based phase corruption in the aperture plane, \cite{Thyagarajan+Carilli2022}  present a geometric understanding of how this invariance manifests itself in the Fourier domain, namely, the image plane. Closure phases are geometrically encoded in the shape, orientation, and size (SOS) of the triangle enclosed by the fringes of a three element interference pattern in the image plane. Such an image from a triad of array elements corresponds to the simple sum of the three sinusoidal fringes in the image plane formed by the corresponding three baseline visibilities in the aperture plane. The characteristic grid patterns formed in such triad images can be seen in examples in \cite{Thyagarajan+Carilli2022}. The important result in their analysis was that, regardless of whether the element-based phase errors in the visibilities are corrected or not, the SOS parameters of the triangle enclosed by the fringes are conserved, with the only degree of freedom being an unknown translation of the grid pattern of triangles due to the element-based phase errors. Thus, the three-fringe interference image from any triad of elements, even if uncalibrated, i.e. uncorrected for element-based phase errors, represents the true sky image that can be reconstructed from that triad, except for an unknown translation. A specific example of this conclusion is shown in Figure 5 of \cite{Thyagarajan+Carilli2022}, where the nature of the triad grid image changes demonstrably if baseline-based (non-closing) phase errors are introduced, but not if the errors are element-based. 

A phase error of $\delta\phi_{ab}$ for a visibility will shift the corresponding fringe in the image plane by $\delta\phi_{ab}/(2\pi|\mathbf{u}_{ab}|)$ (ie. a $2\pi$ phase error shifts a full fringe spacing) in a direction perpendicular to lines of constant phase. When all three visibilities have phase errors, the net shift of the triad can be obtained by the vector sum of the shifts of the individual fringes as described above. If the phase errors can be factorized into element-based phase terms, \cite{Thyagarajan+Carilli2022} show that the shifts for the three fringes formed by a triad of array elements are constrained such that the characteristic grid pattern of the image is preserved, and the full grid pattern simply shifts relative to the uncorrupted image (again, this is not true for baseline-based, or individual fringe, phase errors). As an example, their Figure 7 shows how introducing a phase error into one element  of the triad (which then affects two visibilities in the triad), leads to a shift of the whole pattern along the uncorrupted fringe. Note that the array geometric parameters, meaning fringe spacing and orientation, are always preserved, since they are set by the element positions. Only fringe position is affected by phase errors. 

A straight-forward means of visualizing how SOS conservation works is given in Figure 4 in \cite{Thyagarajan+Carilli2022}. The key point is that, for any three element interferometer, the only possible image corruption due to an element-based phase screen is a tilt of the aperture plane, leading to a shift in the image plane. No higher order decoherence or image blurring is possible, since three points always define a plane parallel to which the wavefronts are coherent. This is not true for an image made with four or more elements, since multiple planes can then be defined, and higher order decoherence (ie. image blurring) occurs. The shift-only property for element-based phase errors of a three-element interferometer is the image-plane equivalent of the invariance of closure phase to such errors in the aperture plane.

Note that we assume only element-based phase errors are relevant (or, in the image plane, image shifts), and that the amplitudes are reasonably determined and stable. 

An important point is that the translation undergone by each three-element interference image is not the same, and it depends on the degree of the element-based calibration required for the interferometer elements in each triad. That is, without accurate knowledge of these triad-idiosyncratic translations, the different three-fringe interferograms will be incoherent with respect to each other and a simple summation will not yield a coherent image unless these translations can be computed and corrected. Conversely, if the effect of the arbitrary translations can be undone, the sum of the three-fringe images is expected to yield the same synthesized image as from standard aperture synthesis after phase calibration. A related point is that shifting individual visibility fringe images based on a model comparison, as opposed to closed triad images, will simply turn the data into the model identically, in an analogous manner to baseline-based aperture plane self-calibration \cite{Perley1999}. 

\subsection{Image-plane Self-Calibration}\label{sec:IPSC}

We emphasize that the key conclusion from SOS conservation is that, if the instrumental or other measurement phase error terms can be factored into element-based phases, then every three fringe image made from a three element array represents a true image of the sky, modulo an unknown overall translation. 

SOS conservation then raises the possibility of an image-plane self-calibration process (IPSC). The process employs a model sky image, plus the set of triad images made from all the three element interferometers that can be defined in the array. These three element images are cross correlated with the model sky image, to derive the unknown, and idiosyncratic, shifts in the two-dimensional sky coordinates due to element-based phase errors, i.e. the peak in the cross correlation corresponds to the required shift for each triad image to restore the true sky position. The shifted triad images are then summed, leading to a coherent image of the sky brightness distribution. Aperture plane self-calibration also employs source model visibilities to which measurements are compared to derive  element complex gains. 

The self-calibration process is iterative, in both the aperture and image plane, where a starting, simple sky brightness model is assumed, from which initial solutions are derived (element gain phases for visibilities, or triad image shifts in the case of image-plane self-cal). A subsequent, improved model is then synthesized from the corrected data, and shifts are re-derived using this new model.

The self-calibration process converges since the problem is over-constrained. For APSC each measurement corresponds to a visibility, hence, there are $N(N-1)/2$ phase measurements from the visibilities, and $N-1$ unknown element-based phase corrections (relative to a reference interferometer element) \cite{Schwab1980, Schwab1981,Cornwell+Fomalont1999,Perley1999}, hence the APSC process is over-constrained by a factor of $\sim N/2$. 

In our current implementation of IPSC, we derive two shifts (RA and Dec), via cross correlation of each measured triad image with the model image. This spatial correlation is done independently for each triad. Each triad is comprised of three measurements (visibilities), so the over-constraint factor is 3/2, independent of how many antennas are in the full array. The analogy in APSC would be performing self-calibration for an $N$-element array, but using only triads of elements independently to derive the element-based phase corrections. In this case, again there are two unknowns (element phases relative to a reference element), and three measurements (visibility phases). In all cases, the astrometry is set by the input model \cite{TMS2017}. 

This process based on cross-correlating the triad images with a model image is justified on the following basis. The three-fringe triad image corresponds to the sum of three spatial Fourier components. In the context of Fourier transforms, the model image is the sum of many Fourier sinusoidal components, which correspond to orthogonal basis functions. In theory, only the three Fourier components in the triad image survive the process of cross-correlating the triad images against the model image (which contains all the Fourier basis components), while the rest of the Fourier components vanish due to orthogonality. The cross-correlation essentially picks up the shifts required for each set of three Fourier components in the different triad images. The summing of appropriately shifted triad images reconstructs all the measured Fourier components present in the data to obtain a dirty image. This process is then verified numerically in the analysis below.

We note that in the low signal-to-noise (S/N) regime per coherence time, as is usually the case in optical interferometry, a common practice is to exploit the fact that closure phase is a `good' observable even on short coherence time scales. This is done by calculating the phase of the vector average of the bispectra over a large number of time frames to build a significant S/N. These closure phases along with separately calibrated visibility amplitudes then undergo a hybrid mapping process to obtain the best-fit model image to the averaged closure phases and visibility amplitudes \cite{Readhead+1988,Nakajima+1989}. In the present context, one can envision fitting the triple fringe patterns with sinusoids (or a direct Fourier transform), to obtain the closure phases and carry out the hybrid mapping process. However, we are presenting a different approach assuming high S/N, in which the self-calibration is performed using triad-based images and image shifts for each coherence time entirely in the image plane.

\subsection{Configuration and Models}\label{sec:config}

\begin{figure*}
\centering
\includegraphics[trim=0in 0in 0in 0in, clip, width=0.7\linewidth]{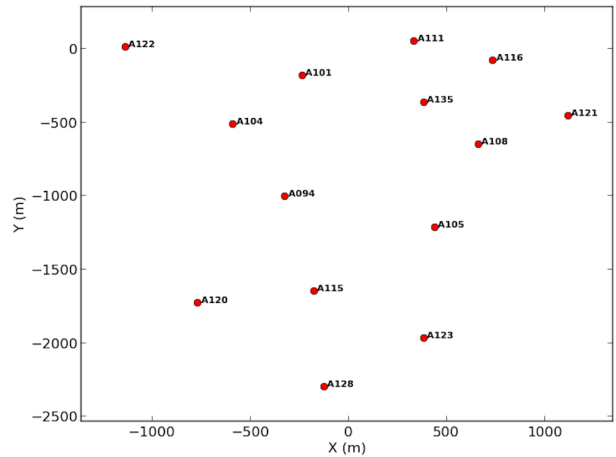}
\caption{Interferometer array layout used for the simulations. The array element distribution is based on the ALMA configuration, with some of the inner element removed to facilitate processing, and potentially to reduce symmetrization artifacts.  
}
\label{fig:array}
\end{figure*}

For this demonstration paper, we perform our imaging and calibration simulations in the context of radio astronomical interferometric measurements, at an observing frequency of 8~GHz. We adopt a configuration from the Atacama Large Millimeter Array (ALMA) set of configurations, with a maximum baseline of 2500~m. We remove inner antennas to avoid very short spacings for the following practical reasons. First, short baselines sample low spatial frequencies (large spatial scales), which then require very large images which greatly increase the processing time. Second, it has been demonstrated that triads with long aspect ratios (one short baseline and two long baselines), have a greater tendency to lead to spurious source symmetrization during the self-calibration process (see Sec~\ref{sec:discussion}; \cite{Linfield1986}). The final configuration has 14 antennas and a spatial dynamic range of six (maximum to minimum baseline length). The configuration is shown in Figure~\ref{fig:array}. A key point is that the results can be scaled to interferometric measurements at any wavelength, using the relative observing wavelength and baseline lengths, as discussed below. 

We adopt a simple source model for our initial tests of the procedure.  We consider two point sources, with a dominant bright source, and a faint companion, with a 100/1 flux ratio. Such a physical situation might correspond to a bright radio AGN with a faint companion, either a jet-knot or binary radio supermassive black hole \cite{Tremblay+2016,Bansal+2017,Perley+2017,Burke-Spolaor2011}, or, in optical interferometry, to a binary star system, or dwarf star -- bright planet ('hot Jupiter') system \cite{Vrijmoet+2022}, although our simulation assumes high signal to noise, while most optical astronomical interferometry is done at low signal to noise. For ease of reference, we refer to the bright source as the `star' and the faint source as the `planet' throughout.  

We also perform tests of the procedure with a 10 times fainter planet (1000/1 flux ratio), and a 10 times brighter planet (10/1 flux ratio), to explore the limits of the process in the context of the procedures adopted.

The source separation is $26^{\prime\prime}$, although, again, the relevant quantity for comparison to real world interferometers is the separation measured in array synthesized beams. For the adopted array and frequency, the naturally weighted point spread function has a FWHM $\simeq 3^{\prime\prime}$, implying a source separation $\simeq 9$ synthesized beams. For illustration, typical near-IR interferometers have resolutions of order 1~mas, implying a scaled source separation for our model of 9~mas. In the context of planet imaging, a 9~mas separation corresponds to 1~AU at a distance of 115~pc (note: 1~AU $=1.5\times 10^{13}$~cm, 1~pc $= 3.1\times 10^{18}$~cm).

The model visibility data set is generated using the \texttt{SIMOBSERVE} task in CASA. We assume a single snap-shot observation in time, and single frequency. We have written a python script that then corrupts the visibility phases on a per antenna basis. We adopt a random distribution of antenna-based phase corruptions in the range of $\pm 10^\circ$ in our primary test case, and $\pm 80^\circ$ in a second case. 

No thermal noise is added, and therefore, these tests represent a high signal to noise scenario, even for the companion. Such an approach is reasonable when making a first theoretical assessment of whether a new calibration technique is viable \cite{Cornwell1987}. We will explore the limitations of noise in the future. Further, it is possible that the image-plane self-calibration procedure presented herein is most relevant in situations where thermal noise is not the dominant limitation, which is often the case in laboratory applications of Fourier imaging, such as X-ray crystallography. 

We also only consider snapshot self-calibration. This approach is the simplest for our initial demonstration of the basic principle. Snapshot self-calibration would be relevant in practice when the phase screen coherence time is short compared to any changes in array geometry, and when the signal-to-noise is high enough to obtain reasonable solutions in a coherence time. This situation is often realized in radio astronomy, and may be common in laboratory, radar, or industrial applications of interferometry \cite{Chau+2008, Perley1999, mitsuhashi_recent_2015}.

\subsection{Implementation Details of IPSC}

For our tests, we generate visibility data as described in Section~\ref{sec:config}. We then Fourier transform the visibility data using the CASA package to generate images, without deconvolution (`dirty images'). Images are made individually for all 364 different three-element triads in the configuration. Hence, each triad image is the linear sum of the three fringes for the three baselines defined by the three elements. In our hypothetical implementation discuss in Section~\ref{sec:discussion}, these triad images would be the starting point of the process. 

We use a pixel of $0.25^{\prime\prime}$, or $\simeq 12$ pixels per FWHM of the synthesized beam. We discuss the effects of pixel size in Section~\ref{sec:comparison}. 

We generate all the triad images of the array, corresponding to 364 triad images for a 14 antenna array, derived from the binomial ‘choose three’ formula for N array elements: number of triads $=N(N-1)(N-2)/6$. Summing these triad-images with equal weight then corresponds to a naturally weighted image of the sky. The summed image is then normalized by 364 to obtain sky brightness, since the synthesized beam peak power response in each triad image is assumed by unity in CASA.  The results for the naturally weighted beam using CASA UV gridding and triad imaging and summing, compared to using APCL in AIPS, are consistent to within 1\%.

We calculate the shift for each triad-image by computing the discrete two-dimensional cross-correlation of the triad-image and a model-image (using the {\tt scipy.signal.correlated2d} function). The position of the maximum of this cross-correlation is the translational shift that needs to be applied to the triad images for a coherent sum. The triad-images are shifted (treating any empty pixels on the field edges due to the shift as zero), and then summed together to get the IPSC image. 

After summing all the triad images, we deconvolve the resulting array synthesized beam using an image-based deconvolution as implemented in the AIPS task APCL, incorporating the appropriate synthesized beam point spread function \cite{Perley1999}. We set CLEAN boxes around the star and/or the star and planet (Section~\ref{sec:tests}), and deconvolve with 3000 iterations and a loop gain of 0.03.

\section{Analysis}

\subsection{Synthesized model image}

\begin{figure*}
\centering
\includegraphics[trim=1.in 2.5in 0in 0in, clip, width=1.1\linewidth]{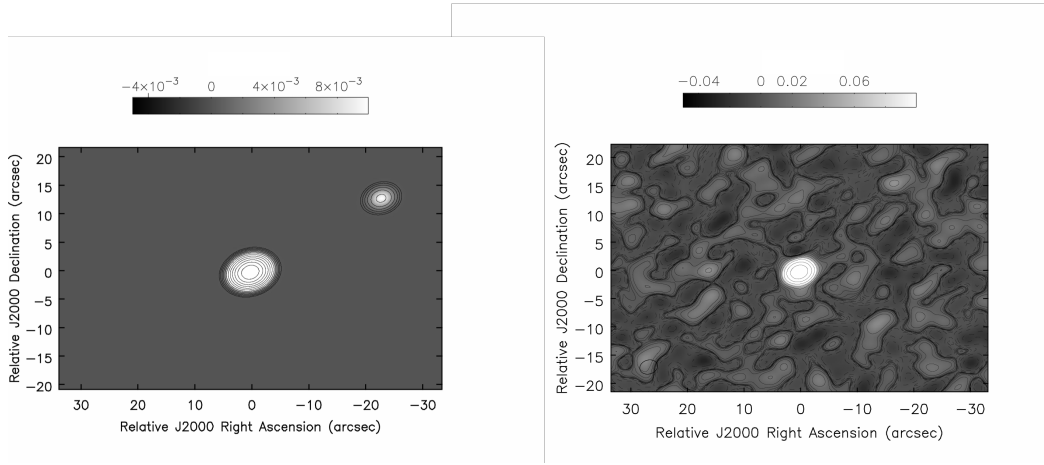}
\caption{Results for the processing on the star and planet field with a flux ratio of 100/1. The star has a flux of 1~Jy in all cases. Left: input model with no phase errors, processed through the interferometric imaging and deconvolution procedures described in Section~\ref{sec:process}. The grey scale runs from -0.01 to 0.01 Jy beam$^{-1}$, and the contour levels are a geometric progression in factor two, starting at 0.25 mJy beam$^{-1}$. Negative surface brightness contours are dashed. The measured off-source rms brightness fluctuations on this image, due strictly to UV-sampling, pixelization and deconvolution, are only 6.4 $\mu$Jy beam$^{-1}$, implying a dynamic range of $1.6\times 10^5$. Right: Same model, but with $10^\circ$ phase errors and no self-calibration. The color scale goes from -0.05 mJy beam$^{-1}$ to 0.1 mJy beam$^{-1}$, and the contours are again a geometric progression in factor two, starting at 0.25 mJy beam$^{-1}$. The rms brightness fluctuations on this image are 9 mJy beam$^{-1}$, implying an image dynamic range of 100. This image shows that the planet cannot be detected with $10^\circ$ phase errors due to residual sidelobe confusion.
}
\label{fig:mod}
\end{figure*}

Figure~\ref{fig:mod} left shows the input model with a 1~Jy star, where 1~Jy $= 10^{-23}$ erg s$^{-1}$ cm$^{-2}$ Hz$^{-1}$, and a 10~mJy planet, both point sources, with no phase errors, but sampled by the interferometer, imaged, and deconvolved, using the processes described in Section~\ref{sec:IPSC}. We show this to define the limits of the image quality set just by the adopted interferometric imaging and deconvolution process. The resulting flux density for the planet, the symmetrized ghost source (see Section~\ref{sec:10d}), and the off-source rms brightness fluctuations on the image, are listed in Table~1. The image shows the planet within 0.6\% of its true flux, no ghost source, to within the rms ($< 1\sigma$), and a dynamic range (star/rms) of $1.6\times 10^5$, nominally set by the UV-sampling, imaging, and deconvolution process. 

Two important points to note. First, the planet at 100/1 ratio cannot be seen in images that are not deconvolved: the rms brightness fluctuations due to the sidelobes of the PSF from the Star are 74~mJy beam$^{-1}$, well above the planet flux, even for uncorrupted data.

Second, the planet is also not seen in images that are deconvolved, but have uncorrected $10^\circ$ phase errors (Figure~\ref{fig:mod}, right). The off-source rms brightness fluctuations in the deconvolved image without self-calibration are 9 mJy beam$^{-1}$, due to residual sidelobes of the synthesized beam due to phase corruption (see Table~1), implying the planet would only be $\sim 1\sigma$. Using equation 13.8 in \cite{Perley1999}, the expected snap-shot dynamic range (peak to rms) for a 14 antenna array with $10^\circ$ random phase errors is $\sim 60$, which is consistent with our results.

\subsection{Self-calibration: 100/1 model}\label{sec:10d}

\begin{figure*}
\centering
\includegraphics[trim=1in 1.3in 0in 0in, clip, width=1.1\linewidth]{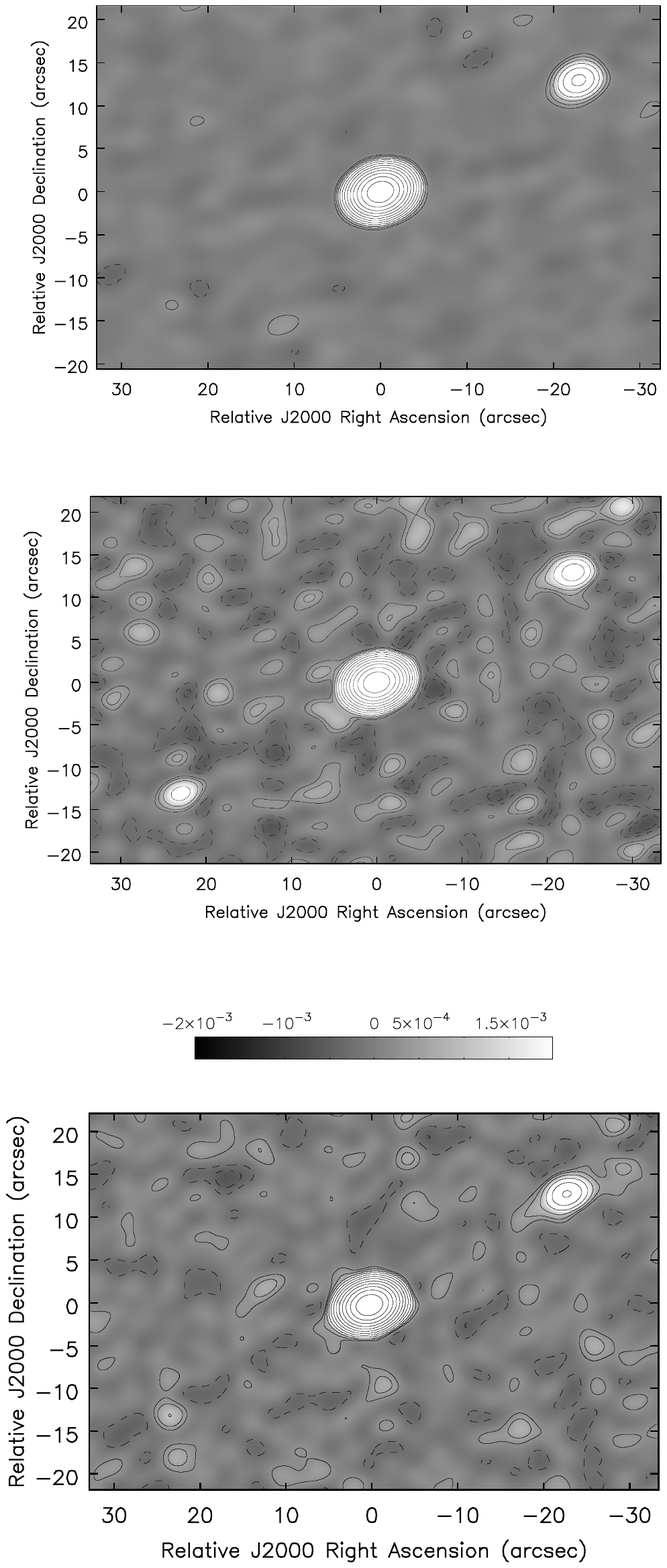}
\caption{Results for the processing on the star and planet field with a flux ratio of 100/1. The star has a flux of 1~Jy in all cases. Upper:  Results for a $10^\circ$ phase error data set, after aperture-plane phase self-calibration using a star-only model. Middle: Same, but using image-plane self-calibration with a star-only model. Lower: Same, but using image-plane self-calibration with a star and planet model. In all cases, the contour levels are a geometric progression in factor two, starting at 0.25 mJy beam$^{-1}$. Negative surface brightness contours are dashed. The grey scale goes from -0.002 mJy beam$^{-1}$ to 0.002 mJy beam$^{-1}$. 
}
\label{fig:10d}
\end{figure*}

We start with the results of self-calibration in the aperture plane (APSC), as a reference point. Figure~\ref{fig:10d} upper panel shows the resulting image for the $10^\circ$ phase error uv-dataset after phase-only self-calibration with a star-only, point source model using the task \texttt{GAINCAL} in CASA. Table~1 lists the planet and ghost source flux density, and the off-source rms brightness fluctuations. The planet is 3\% fainter than expected. The rms is a factor 14 times higher than in the model image with no errors. At the position of the symmetrized ghost source, there is  a $2\sigma$ feature that could be residual symmetrization even with aperture plane self-calibration. 

The lower two frames in Figure~\ref{fig:10d} show the results after IPSC, and the relevant image quantities are listed in Table~1. The center image is after a single iteration using a star-only, point-source model. In this case, the planet is clearly visible at $\sim 10\sigma$, but the flux density is only 63\% of the expected value. 

Importantly, a ghost source appears symmetrically reflected across the star position to the southeast, with a flux about 26\% of the model planet flux. The rms is about three times larger than for the aperture plane self-calibration.

We then iterate, increasing the planet flux in the model based on the results from the previous self-calibration cycle. The planet gets stronger systematically, while the ghost source fades. The final result, where the input model is close to the true model, is shown in the lower frame in Figure~\ref{fig:10d}, and image values listed in Table~1.

The IPSC is converging, with a planet flux $\sim 90$\% of its true value,
and off-source image brightness fluctuations that are a factor two larger than the APSC result. The ghost source is reduced to $\sim 3\sigma$, or 6\% of the model planet flux.

Note, we do not include the ghost source in the iterative self-calibration model for a number of reasons. First, a perfectly symmetric source distribution around the central star seems physically implausible. And second, self-calibration has historically been known to generate 'symmetrized' artifacts around bright sources. We discuss this in Section~\ref{sec:comparison}.

\subsection{Further tests}\label{sec:tests}

We have repeated the IPSC process using the 100/1 model, and much larger initial random antenna-based phase errors over the range of $\pm 80^\circ$. Table~1 lists the resulting image parameters after iteration of the self-calibration process, starting with a point source model, then to a model including the planet. The planet can be easily identified as the strongest source (besides the star) in the first iteration using a star-only sky model. The process converges to a final image  comparable in quality to the $10^\circ$ phase error case (Table~1).  In this case, the larger phase errors led to significant decoherence of the star itself, with a peak surface brightness of only 0.71 Jy beam$^{-1}$ without any self-calibration. After IPSC, the star is recovered at full coherence, with 1.00 Jy beam$^{-1}$. 

We also perform IPSC of the 100/1 data set with the $10^\circ$ errors, using an incorrect starting model, with the planet moved to the wrong position. The result is comparable to a single iteration of star-only model self-calibration (see Table~1). A negative source does appear at the wrong planet position at about 13\% of the flux of the planet (Table~1).

\begin{figure*}
\centering
\includegraphics[trim=.7in 2.5in 0in 0in, clip, width=1.1\linewidth]{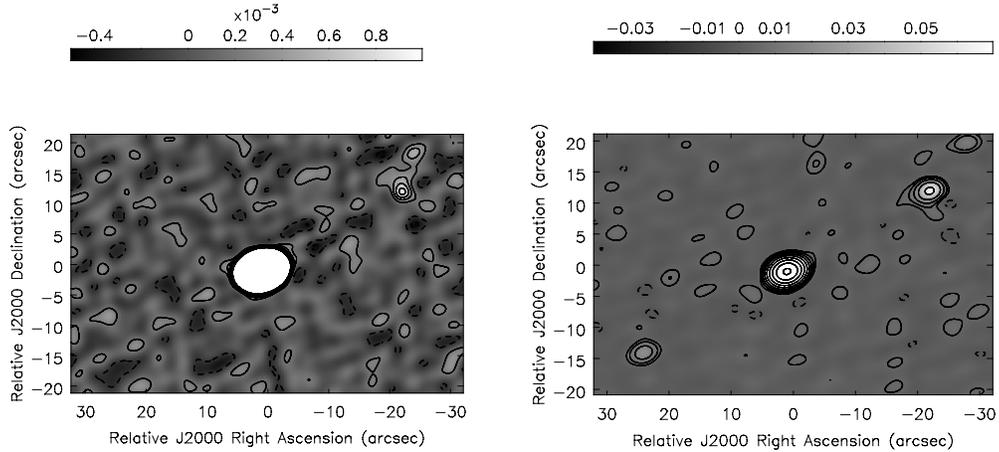}
\caption{Left: results for a 1000/1 star to planet ratio and a $10^\circ$ phase error data set, after image-plane self-calibration with a star-only model. The contour levels are linear, starting at $0.2$~mJy~beam$^{-1}$, in increments of $0.2$~mJy~beam$^{-1}$. Negative contours are dashed. The color scale runs from -0.5 to 1~mJy~beam$^{-1}$. Right: results for a 10/1 star to planet ratio and $10^\circ$ phase errors, after IPSC with a star-only model. The contour levels are a geometric progression in factor two, starting at $3.5$~mJy~beam$^{-1}$, in increments of $0.2$~mJy~beam$^{-1}$. The color scale runs from -40 to 70~mJy~beam$^{-1}$. 
}
\label{fig:brightfaint}
\end{figure*}

Next, we investigate a model sky with the planet flux decreased to 1 mJy, implying a 1000/1 star-planet flux ratio. We then perform star-only IPSC, and the planet does appear as the strongest source in the field (besides the star), with a flux 72\% of the true value, and a significance of $\simeq 4.8\sigma$ (see Table~1), even if we do not include the planet in the CLEAN boxes ('1 box' example in Table~1; Figure~\ref{fig:brightfaint}). The ghost source is not seen, to within the rms level. Including the planet in the CLEAN boxes does not change the rms level, but increases the planet flux to $\simeq 6.2\sigma$. It is clear that the fainter planet is not significant enough to alter the IPSC shift solutions. Interestingly, if the ghost source for a star-only self-calibration had the same ratio as the 100/1 case, we would expect to see it at the $\sim 2\sigma$ level in the image. The lack of such a $2\sigma$ ghost suggests that, for a highly point-source dominated model, symmetrization may be less relevant.

Lastly, we investigate a model sky with the planet increased to 100 mJy, implying a 10/1 star/planet flux ratio. We perform star-only self-calibration, and the planet is clearly evident as the strongest source in the field other than the star, with a flux 65\% of its true value (see Table~1). The ghost source is seen at 21\% of the model planet flux, comparable to the ratio seen for the 100/1 model. The off-source rms brightness fluctuations on this image are higher than for the fainter planet models, likely due to the fact that the self-calibration solutions using the star-only self-cal model are worse, due to the much brighter planet. Iterating to a final star plus planet self-calibration model leads to a planet of 88\% of its true value. The ghost is reduced to 7\% of the true planet flux, again, comparable to the final result for the 100/1 star/planet model, and the rms is reduced by a factor of $\simeq 3$. 

\subsection{Comparison of APSC and IPSC}\label{sec:comparison}

IPSC process does not converge as well as the APSC for large-$N$ arrays, as can be seen from the results shown in Table~1 and Figure~\ref{fig:10d}. The measured image rms for APSC is a factor two lower than for IPSC, and the planet itself is 99\% its true strength with just one iteration of APSC using a point source model, while for IPSC it only reaches 90\% after iteration to a star-planet model. 

The results in the IPSC are partially dictated by pixelization limitations. For instance, the pixel size we employed was $1/12~ \times$ FWHM of the synthesized beam. For a shift relative to a pixel center, this implies a maximum error of ~$\sim ~1/24~ \times 360^o = 15^o$, which is of the same magnitude as the imposed errors in the simulation process. Of course, this maximum error can only manifest itself on the longer baselines, which provide the higher resolution distinct in the synthesized beam. On the triad images incorporating shorter baselines, with larger fringe frequencies but the same pixel size, the numerical errors are proportionately smaller.  

As an example, we have run tests using a larger pixel size ($0.5"$ vs. $0.25"$ or $1/6~\times $ FWHM vs. $1/12~ \times$ FWHM of the synthesized beam), and a smaller pixel size of $0.16"$. For the larger pixels, the resulting image rms increases by a factor of 1.35, the planet is weaker by 6\%, and the ghost source is stronger by a factor of two. For the smaller pixels, the off-source rms is lower (0.16 mJy vs. 0.20 mJy), while the star and planet fluxes are the same to better than $1\sigma$. The result implies that, as the phase errors due to pixelization drop below the imposed phase errors, further improvement is marginal. These results are listed in Table~\ref{table1}. 

The more fundamental limitation of IPSC, as implemented herein, is that, for larger-$N$ arrays, APSC is more over-constrained than the IPSC, as discussed in Section~\ref{sec:IPSC}. We are investigating the possibility of determining globally optimized shifts using the information from all the triad images at once, since all the triad image shifts (a total of $N(N-1)(N-2)/6$ of which $(N-1)(N-2)/2$ are linearly independent) are postulated to be derived from only $N-1$ element-based phase errors. With mutually shared elements in the different triads being taken into account in the global optimization, it may increase the over-constraint factor. However, in optical interferometry, the incident light will have to be split into larger number of beams which may have a detrimental effect overall if the incident signal to noise is low. A detailed investigation needs to be undertaken and is outside the scope of this proof-of-concept paper. 

Another difference in the results from APSC and IPSC is the generation of the `ghost source', symmetrically placed across the star from the planet. For IPSC with just a star model, the symmetrized source is 2.6~mJy, while for APSC the symmetrized source was only 0.2~mJy. Further iteration of IPSC reduces the ghost to 0.6~mJy.

Source symmetrization due to self-calibration is a well known phenomenon, in particular for models/images with a dominant point source \cite{Cornwell+Fomalont1999,Linfield1986}. A related symmetrization was also discussed as part of the `uniqueness problem', in the original work on phase retrieval using image intensity measurements by \cite{Gerchberg+Saxton1972,Millane1990}, particularly in the context of X-ray crystallography and electron microscopy.  \cite{Cornwell+Fomalont1999} point out: `... if the number of elements is sufficiently small, then the corrected phases will be significantly biased toward zero. As a consequence, after one iteration of self-calibration some features in the image will be seen reflected relative to the point-like component.' \cite{Linfield1986} showed that this problem can be exacerbated for highly acute closure triangles, i.e. one very short baseline and two long baselines, as may be the case in e.g. space VLBI. In this case, for the closure phase sum, the visibility phases due to source structure on the two long baselines cancel, and the closure phase is then approximately the visibility phase due to structure on the short baseline, which, in the case of very short baselines, is close to zero. Hence, we have removed the antennas that made the shortest baselines from the array layout in this analysis. 

Still, we have found that the IPSC process leads to greater source symmetrization than the APSC for this layout, although subsequent iteration reduces the symmetrization substantially. We feel it likely that the worse symmetrization in the IPSC process again relates to the less-highly constrained solutions relative to APSC.

\section{Discussion} \label{sec:discussion}

We present a new method to generate phase-coherent images of a source with an interferometer, based on the Shape-Orientation-Size (SOS) conservation principle, in situations where interferometric phase errors can be factorized into element-based terms. The self-calibration method is implemented in the image domain, without resort to antenna-based complex voltage gains. The image plane self-calibration process relies on SOS conservation for imaging with triads of array elements, and entails cross-correlation of these triad images with models of source brightness, to derive the unknown triad image shifts due to element-based phase errors. After correcting for these shifts and summing, the source brightness distribution is recovered at diffraction limited resolution, at least in the case of relatively simple source models.

We have shown that, in the case of a simple source model consisting of two point sources with a flux ratio of 100/1, the process allows for identification of the two source components, and converges on a final image that is a reasonable representation of the brightness distribution. We have verified this conclusion using both small ($\pm10^\circ$) and large ($\pm 80^\circ$) phase errors, an incorrect starting model, and fainter and brighter planets. In all cases, we have done our tests in the high signal-to-noise regime, to test the basic principle. 

We have also found the aperture plane self-calibration converges more quickly, and to a better final image, than the image-plane process, at least for the case studies herein, due to the higher level of over-constraints for a large-$N$ array. Hence, any consideration of IPSC, as currently implemented, would only be relevant in the context of small-$N$ arrays. There are numerous examples of $N \le 4$ interferometers in optical and near-IR astronomy, on the ground and in space \cite{Gravity+2017, Quanz+2021, Hansen+Ireland2020, LeDuigou+2017}. However, such interferometers typically operate in the low S/N regime per coherence time, and hence may not be immediately relevant in those cases. A possible application in the high signal to noise regime may be in laboratory or industrial interferometers, where the signal can be strong.

For reference, we consider a hypothetical device based on IPSC. Such a device would involve a direct imaging beam combiner (eg. Fizeau beam combination \cite{Lardiere+2007}), but combining signals for each 3-element sub-array separately. The SOS conservation principle then implies these triad images are true images of the sky, with unknown shifts due to piston phase errors. The IPSC process can then be employed to derive the offsets for each coherence time, and summing then yields a corrected image of the source. In this case, adaptive optics to correct for the phase screen are not required. The delay errors are corrected  via the IPSC process. As an example, a 4-element array would require three beam splits for the signal from each element, and four CCD detectors to record the different triad images at any give time. 

We re-emphasize that the IPSC process considered herein is in the context of high signal-to-noise, which is not the case for many astronomical interferometers. However, the technique is also applicable in some laboratory, radar, and industrial applications where the signal-to-noise ratio may be high, but wavefront phase corruption remains an issue, and hence image reconstruction requires phase correction \cite{Chau+2008} . In the labaratory, one potential application is to synchrotron radiation interferometry \cite{mitsuhashi_recent_2015, torino16}, a technique for measuring the synchrotron beam used during calibration and commissioning of synchrotron facilities. Here IPSC would allow correction of wavefront errors due to optics and in-air propagation and allow a more straightforward reconstruction of an image of the synchrotron beam. 

A possible advantage of IPSC is that simple processing requirements mean that very high frame rates can be processed in real time to reconstruct the images. High frame rates are common in the laboratory, where bright artificial radiation sources can be used. In such a lab setting a mechanical or a microoptoelectromechanical system may be used to rapidly reconfigure the apertures in order to fill-out the $uv$-plane for a small-$N$ array. For example, an addressable array of very small mirrors (a digital micromirror device, \cite{hornbeck_spatial_1991,dudley_emerging_2003}) may be used to select three apertures from a larger aperture plane at reasonable rates leading to possibility of good quality diffraction-limited imaging on sub-second time scales.

This paper is the first investigation of a new technique in interferometric calibration, and we have adopted a simple approach and model for initial demonstration. While the process produces coherent images of a source, real-world applications where IPSC is competitive remain to be explored. Still, as a technique, it provides insight into the nature of closure phase and self-calibration. 

{\bf Acknowledgments.} The National Radio Astronomy Observatory is a facility of the National Science Foundation operated under cooperative agreement by Associated Universities, Inc.. Software: Astronomical Image Processing System (AIPS) \cite{Greisen2003}. Common Astronomical Software Applications (CASA) \cite{casa:2017}. We thank Preshanth Jagannathan, Chris Haniff, and the referee for useful comments. 

{\bf Disclosure Statement:} The authors declare no conflicts of interest.

{\bf Data availability:} The simulated data underlying the results presented in this paper are not publicly available at this time but may be obtained from the authors upon reasonable request.

\clearpage
\newpage

\begin{table}
\centering
\caption{Observed Properties}
\begin{tabular}{lcccc} 
\hline
 Data & Planet & Ghost & rms \\
 ~ & mJy$^a$ & mJy & mJy beam$^{-1}$ \\
 \hline
Model: star = 1000~mJy, planet = 10~mJy, pix $= 0.25''$ & 9.94  &  0.0052 & 0.0064 \\
$10^\circ$ errors; No Self-cal &  5.8  &  0.9 &  9 \\
$10^\circ$ errors; aperture self-cal star only &  9.70  &  0.20 &  0.089 \\
$10^\circ$ errors; image self-cal star only &   6.26 &   2.63 &  0.25 \\
$10^\circ$ errors; image self-cal star and planet &  8.97 &   0.62 &  0.20 \\
$80^\circ$ errors; image self-cal star and planet &  8.61 &  0.84 &  0.18 \\
$10^\circ$ errors; image self-cal wrong planet &  6.70  &  2.20 & 0.30 \\
\hline
$10^\circ$ errors; image self-cal star and planet, pix = $0.5''$ &  8.5  &  1.3 & 0.27 \\
$10^\circ$ errors; image self-cal star and planet, pix = $0.16''$ &  8.85  & 0.75 & 0.16 \\
\hline 
Model: star = 1000~mJy, planet = 1~mJy & & & \\
$10^\circ$ errors; image self-cal star only, 1 box & 0.72 &  0.025 & 0.15 \\
$10^\circ$ errors; image self-cal star only, 2 boxes & 0.88  &  0.096 & 0.14 \\
$10^\circ$ errors; image self-cal star and planet, 2 boxes &  0.84 & 0.15  & 0.15 \\
\hline 
Model: star = 1000~mJy, planet = 100~mJy & & & \\
$10^\circ$ errors; image self-cal star only & 65.0 &  21.0 & 1.9 \\
$10^\circ$ errors; image self-cal star + planet & 88.1 &  6.9 & 0.69 \\
\hline
\end{tabular}
\label{table1}
\end{table}
$^a$ We adopt units of mJy and not mJy beam$^{-1}$ since the sources appear unresolved in all cases, as per the input model. 

\bibliography{refs}



\end{document}